# ФИЗИЧЕСКИЕ АСПЕКТЫ ПРОЦЕССОВ САМООРГАНИЗАЦИИ В КОМПОЗИТАХ

## 1. МОДЕЛИРОВАНИЕ ПЕРКОЛЯЦИОННЫХ КЛАСТЕРОВ ФАЗ И ВНУТРЕННИХ ГРАНИЦ

Герега А.Н.

*Одесская государственная академия строительства и архитектуры, Украина*

**РЕЗЮМЕ**

Рассмотрен комплекс перколяционных моделей кластерных систем композитов. В моделях методом Монте-Карло получены параметры кластеров вещества и внутренних границ, изучена возможность влияния на свойства композиционных материалов на ранних стадиях формирования структуры. Исследованы модели технологических трещин, изучены особенности эволюции кластеров вещества с трансформирующимися элементами.

Аналитически решена континуальная перколяционная задача на ковре Серпинского с гибридной разветвленностью. Предложена система имитационного моделирования алгоритмов построения и исследования кластерных структур в материалах.

**Ключевые слова:** моделирование структуры; внутренние границы; ковер Серпинского; фрактальные структуры; перколяция; критические индексы; моделирование алгоритмов

# PHYSICAL ASPECTS OF THE PROCESSES SELF-ORGANIZATION IN COMPOSITES

## 1. SIMULATION OF PERCOLATION CLUSTERS PHASES AND INNER BOUNDARIES

Herega A.N.

*Odessa State Academy of Civil Engineering and Architecture, Ukraine*

**SUMMARY**

In paper submit the percolation models of cluster systems composites. In the models by Monte-Carlo method obtained the parameters of clusters substances and internal borders, studied the possibility of influence on the properties of composite materials in the early stages of structure formation. Authors investigated models of technological cracks, and peculiarities of the evolution of clusters of matter with transforming elements.

Analytically solved the continual percolation problem on the Sierpinski carpet with hybrid branching. We proposed system simulation algorithms for the construction and study of cluster structures in materials.

**Key words:** simulation of structure; inner boundaries; Sierpinski carpet; fractal structures; percolation; critical exponents; modeling of algorithms

# 1. ВВЕДЕНИЕ

Раздел теории вероятностей, имеющий многообразные приложения в естественных и инженерных науках, – перколяционная теория – на протяжении более полувека изучает особенности возникновения и эволюции, а также свойства кластеров в матрицах произвольной структуры и размерности [1-3].

Образующиеся перколяционные кластеры кардинально модифицируют материал: в нем происходит структурный фазовый переход, скачкообразно возрастает корреляционная длина, меняется симметрия объекта и другие параметры, что приводит к изменению физико-химических и механических характеристик физических тел. Наблюдающийся на протяжении последних десятилетий устойчивый интерес специалистов к перколяционным кластерам легко объясняется очевидной важностью изучения критических явлений: вблизи точки фазового перехода, благодаря большим размерам кластеров, геометрия системы фактически не зависит от вида вещества, а обладает универсальными свойствами, присущими системам различной природы [1-3]. Эти кластеры существенно видоизменяют процессы проводимости, влияют на кинетику химических реакций, определяют механическую прочность и коррозийную устойчивость, приводят к аномальной диффузии и другим явлениям. Поэтому в перколяционных исследованиях обычно одновременно изучается и кластерная система физического тела, и её влияние на объект в целом.

Перколяционная теория предоставляет адекватный математический аппарат и возможность физического описания явлений, причиной которых является возникновение связных (квазисвязных) областей, имеющих характерные размеры тела, – перколяционных кластеров произвольной природы: фаз, дефектов, границ раздела и др.

Целью статьи является описание моделей, созданных для изучения зависимости структуры и свойств перколяционных кластеров от характерных размеров и формы, количества и распределение по размерам, а также характера взаимодействие малых кластеров. Статья содержит результаты исследования возможностей управления свойствами композиционных материалов на ранней стадии их генезиса.

# 2. КОМПЛЕКСНАЯ ПЕРКОЛЯЦИОННАЯ МОДЕЛЬ СТРУКТУРЫ МАТЕРИАЛА

«Перколяционный бум» семидесятых годов прошлого века показал, что теория протекания имеет такое количество модификаций алгоритмов построения связных областей, что «несть им числа» [1-5].

Для изучения структуры промежуточной асимптотики, строения и свойств перколяционных кластеров на поверхности и в объеме твердых тел разработана полифункциональная перколяционная модель [4-7]. Модель предназначена для изучения перколяционных кластеров, построенных методом Монте-Карло по широкому спектру алгоритмов, что позволяет использовать ее для изучения разнообразных явлений и процессов, таких как прыжковая проводимость в твердых телах, критические явления в кинетике твердотельных химических реакций, аномальная диффузия, деструкция материалов и другие.

В зависимости от цели исследования модельные кластеры, в том числе перколяционные, играют роль объектов различной физической природы: объемов, занимаемых некоторой фазой вещества, внутренних границ, микроскопических дефектов различной природы, трещин и других.

В модели предусмотрена возможность расчета физических и геометрических характеристик кластерных систем: радиуса гирации, степени анизотропии, корреляционной длины, распределения кластеров по размерам, фрактальной, информационной и корреляционных размерностей из спектра Реньи, лакунарности и других (см. Приложение), а также статистическая обработка результатов экспериментов. В основу компьютерной реализации модели положен объектно-ориентированный подход, что позволяет легко модифицировать ее возможности, используя новые классы, свойства, методы.

### 2.1. Плоские модели

Построение перколяционного кластера проводится методом Монте-Карло на квадратном поле размером $370 \times 370$ условных единиц длины. Частицы, из которых строятся кластеры, имеют форму кругов (для некоторых задач – окружностей); в каждом модельном эксперименте принимается решение о характере распределения частиц по размерам: он либо фиксируются, либо выбирается из дискретного нормального распределения возможных значений. Обычно, в модельных экспериментах сторона поля превышает размеры частиц в 40-50 раз.

В качестве алгоритма роста кластеров в модели выбран путь последовательного наращивания координационных сфер (рис.1). Управляющими параметрами модели являются количество центров кластерообразования, отношение между диаметром частицы и стороной поля (относительный размер единичной частицы), условие возникновения связности между элементами, а также соотношения между размерами частиц, образующими кластеры. На рис. 2 показаны кластеры, сформированные при различных величинах управляющих параметров.

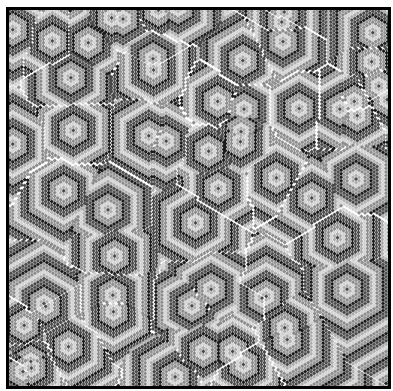

Рис. 1. Координационные сферы, наращиваемые на случайных центрах, частицами одинакового размера.

В заполнении сфер существенную роль играет генератор случайных чисел с равномерным распределением. Сначала с его помощью выбираются координаты центров кластерообразования, затем – тот из центров, в котором будет происходить очередной акт заполнения, и, наконец, генератор указывает место в заполняемой координационной сфере, где будет расположена очередная частица растущего кластера. В случае если размер элементов не фиксирован, он выбирается из массива размеров кругов (больший из которых равен диаметру из эксперимента с одинаковыми кругами) с помощью генератора с нормальным распределением. Генерация прекращается, когда сформирован перколяционный кластер частиц.

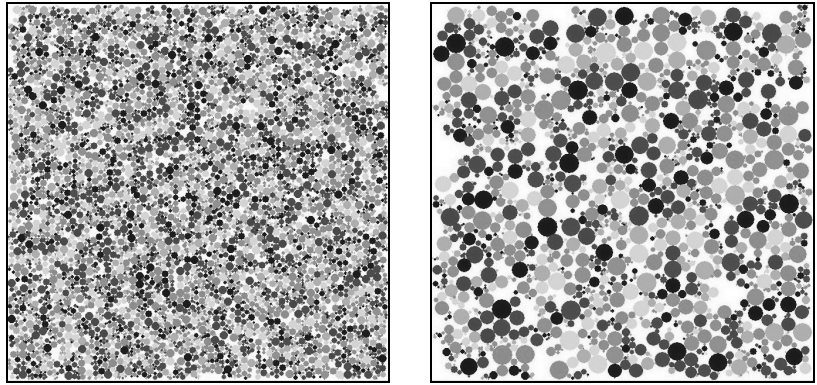

Рис. 2. Кластеры из частиц различной величины, полученные при варьировании значений управляющих параметров.

### 2.1.1. Кластеры фаз и внутренних границ

В модели фактически формируются две кластерные системы – частиц и внутренних границ или пустот, которые являются фоном одна для другой, как в мозаиках Маурица Эшера (рис.3). Что считать кластерами зависит от цели исследования, также как, например, восприятие зрительных образов во много предопределяется установкой наблюдателя.

На рис. 4 изображен перколяционный кластер границ раздела.

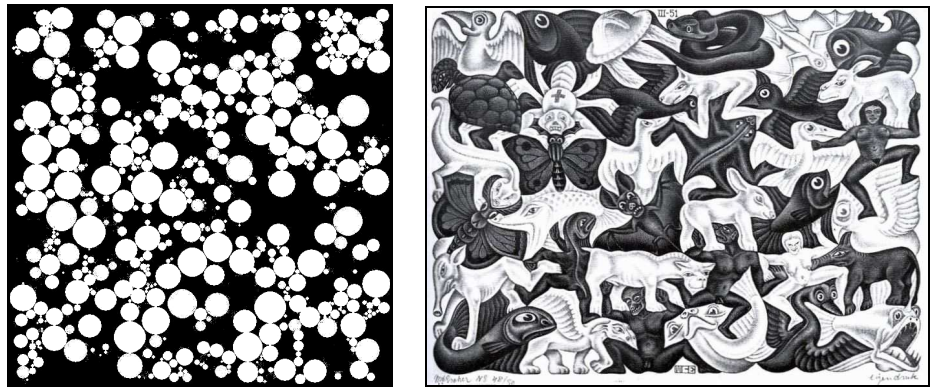

Рис. 3. Модельные кластеры и гравюра М. Эшера «Мозаика I».

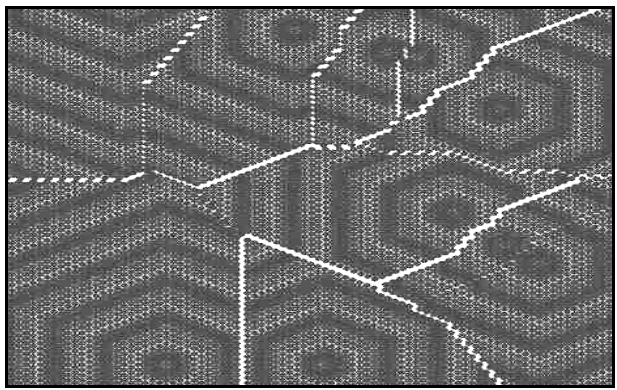

Рис. 4. Перколяционный кластер внутренних границ.

В табл. 1 показаны результаты статистической обработки значений параметров конечных и перколяционных кластеров. Представлены только величины, значения которых позволяют получить представление о «физике и геометрии» кластеров. Характерная относительная погрешность расчетов – (10 ± 3)%. Используемые расчетные формулы описаны в Приложении.

По данным таблицы видно, что дисперсный состав частиц существенно влияет на свойства кластерных систем материала. Например, в случае если частицы имеют произвольный диаметр (при описанном выше условии, что максимальное значение диаметра из гауссова набора равно величине фиксированного диаметра из другой реализации модели), мощность кластерной системы частиц, примерно, в 1.5 раза больше, чем при фиксированном. Корреляционная длина – в первом приближении она соответствует характерному размеру конечных кластеров частиц – в среднем, примерно, в 3 раза меньше в случае произвольного диаметра, как и радиус гирации, величина которого особенно существенна на стадии формирования материала.

Характерны прочерки в правой колонке таблицы, где должны находиться характеристики бесконечного (перколяционного) кластера внутренних границ: эти параметры получены с относительной погрешностью, превышающей 15%. Это связано с размерностью пространства, в котором происходит перколяционная «игра»: в такой постановке задачи бесконечный кластер границ возникает чрезвычайно редко, типичным является его отсутствие.

### 2.1.2. Моделирование кластеров с пониженным порогом протекания

Полученные результаты показывают, что в предложенной модели кластеры внутренних границ тяготеет к известному классу задач теории протекания, имеющих нулевой перколяционный порог [8].

Авторы [8] считают, что их модель предоставляет полезное описание порового пространства скальных пород. Кроме того, в работе показано, что в перколяционной кластерной системе (например, в полупроводнике) с определенным уровнем протекания, для частиц, обладающих соответствующей энергией, несущественно, насколько мала объемная доля проводящей фазы. Это обстоятельство вызывает ассоциацию с лодкой в заливе с рифами: чтобы пройти через акваторию, лодке достаточен слой воды над рифами, чуть превышающей ее осадку.

В нашей модели возможны несколько вариантов прямого понижения перколяционного порога системы. Во-первых, изменение одного из управляющих параметров модели – расстояния, на котором элементы являются соединенными, – что, естественно, приводит к уменьшению мощности перколяционного кластера с ростом допустимой дистанции. Среди известных задач, решаемых в рамках такой модели, определение электропроводности случайных сеток, задачи прыжковой проводимости, спонтанной намагниченности в ферромагнетике и другие [3].

Другие варианты связаны с возможностями существенной модификации структурных элементов кластеров. Их две: первая – тривиальная, приводящая к предсказуемым результатам, это замена кругов, используемых в модели, на окружности; вторая – использование в качестве структурных элементов либо треугольника или квадрата Серпинского, либо снежинку Коха – предфракталы определенного поколения с габаритами в десятки раз меньшими характерных размеров перколяционного поля. Это позволяет эффективно изменять такие параметры как связность области, корреляционную длину, степень заполнения пространства, лакунарность и др. [5].

Табл. 1.
Некоторые характерные параметры кластерной системы частиц и внутренних границ на поверхности тела.

| Параметры | Фиксированный диаметр | Набор диаметров |
|---|---|---|
| Мощность кластерной системы частиц | $(397 \pm 20) \cdot 10^{-3}$ (5.0%) | $(581 \pm 54) \cdot 10^{-3}$ (9.3%) |
| Мощность наибольшего конечного кластера внутренних границ (НКК ВГ) | $(29.3 \pm 2.8) \cdot 10^{-3}$ (9.6%) | $(20 \pm 1.8) \cdot 10^{-3}$ (11.1%) |
| Мощность перколяционного кластера внутренних границ (ПК ВГ) | $(38.4 \pm 3) \cdot 10^{-3}$ (7.8%) | – |
| Радиус-вектор центра масс наибольшего конечного кластера частиц (НККЧ) | $286.47 \pm 26.82$ (9.4%) | $315.38 \pm 32.06$ (10.2%) |
| Радиус-вектор центра масс ПК ВГ | $288.14 \pm 23.38$ (8.1%) | – |
| Радиус гирации НКК ВГ | $93.08 \pm 9.99$ (10.7%) | $36.05 \pm 3.44$ (9.5%) |
| Радиус гирации ПК ВГ | $116.74 \pm 14.03$ (12.0%) | – |
| Среднее значение радиуса гирации конечных кластеров частиц (ККЧ) | $84.79 \pm 9.65$ (11.4%) | $25.28 \pm 2.61$ (10.3%) |
| Средняя корреляционная длина кластеров частиц | $149.17 \pm 13.68$ (9.2%) | $44.07 \pm 4.73$ (10.7%) |
| Отношение между массой ПК ВГ и радиусом гирации $b = S/R^D$ | $10.41 \pm 1.05$ (10.1%) | – |
| Отношение между массой НККЧ и радиусом гирации $b = S/R^D$ | $10.13 \pm 1.06$ (10.5%) | $9.95 \pm 1.21$ (12.2%) |
| Фрактальная размерность ПК ВГ | $1.32 \pm 0.09$ (6.8%) | – |
| Индекс роста мощности ККЧ | $2.54 \pm 0.16$ (6.3%) | $2.34 \pm 0.12$ (5.1%) |

### 2.1.3. Перколяционные кластеры с трансформирующимися элементами

Еще одна модифицированная перколяционная задача, которая решается в комплексной модели, позволяет описать генезис кластерных структур с трансформирующимися элементами [9].

В задаче, также решаемой методом Монте-Карло, элементы являются соединенными в случае контакта; если они перекрываются, то область наложения считается модифицированной. Предусмотрены четыре типа модифицированных элементов. В модели имеется возможность проследить образование перколяционных кластеров из всевозможных сочетаний элементов различных типов (рис. 4) и рассчитать их параметры.

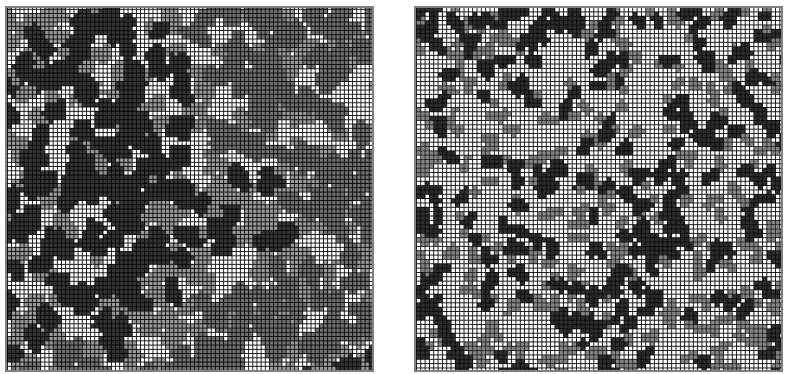

Рис. 4. Кластерные системы с элементами различной степени разупорядочения.

### 2.2. Трёхмерные модели

Объёмная задача реализована в кубе, содержащем $10^6$ ячеек, и не нуждается в отдельном описании: двухмерная задача легко обобщается. На рис. 5 показана характерная реализация перколяционных кластеров в трёхмерной модели: слева – кластер из шаров произвольного диаметра, справа – пустот (внутренних границ) для наглядности в пустом поле.

Результаты модельных экспериментов представлены в табл. 2 и 3.

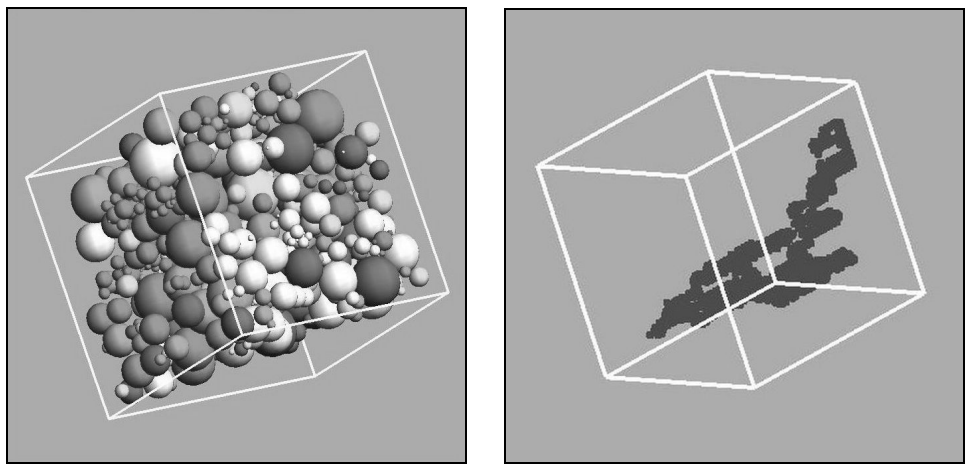

Рис. 5. Кластерные системы объёмной модели.

Табл. 2.

Параметры перколяционных кластеров
частиц и внутренних границ в объёме тела.

| Параметры | Фиксированный диаметр | Набор диаметров |
|---|---|---|
| Мощность кластерной системы частиц | $(440 \pm 50) \cdot 10^{-3}$ (11.4%) | $(636.9 \pm 59) \cdot 10^{-3}$ (9.3%) |
| Мощность ПК ВГ | $(2600 \pm 200) \cdot 10^{-6}$ (7.7%) | $(800 \pm 60) \cdot 10^{-6}$ (7.5%) |
| Радиус-вектор центра масс ПКЧ | $40.73 \pm 4.48$ (10.9%) | $51.57 \pm 5.34$ (10.4%) |
| Отношение между массой ПКЧ и радиусом гирации $b = S/R^D$ | $12.59 \pm 1.35$ (10.7%) | $6.46 \pm 0.65$ (10.1%) |
| Фрактальная размерность ПК ВГ | $1.49 \pm 0.14$ (9.4%) | $1.39 \pm 0.12$ (8.6%) |
| Радиус гирации ПК ВГ | $29.49 \pm 3.45$ (11.7%) | $29.63 \pm 1.66$ (5.6%) |
| Степень анизотропии ПКЧ $A_{xy}$ | $10.64 \pm 1.43$ (13.4 %) | $10.02 \pm 1.35$ (13.5 %) |
| Степень анизотропии ПКЧ $A_{xz}$ | $20.73 \pm 1.71$ (8.3 %) | $26.39 \pm 3.25$ (12.3 %) |
| Степень анизотропии ПКЧ $A_{yz}$ | $4.89 \pm 0.65$ (13.3 %) | $4.86 \pm 0.62$ (12.8 %) |

### 3. ПРЯМОЕ МОДЕЛИРОВАНИЕ ПЕРКОЛЯЦИОННЫХ КЛАСТЕРОВ ТЕХНОЛОГИЧЕСКИХ ТРЕЩИН И ВНУТРЕННИХ ГРАНИЦ

В феноменологической модели возникновения и развития технологических трещин и внутренних границ в объеме тела методом Монте-Карло решается континуальная перколяционная задача [10].

Построение кластерной системы проводится в кубе размером $10^6$ условных единиц длины. Структурные элементы перколяционного кластера – треки – являются аналогами малых технологических трещин. Длина треков может принимать значения 5, 10, 15 единиц длины, а вертикальная и горизонтальная составляющие угла отклонения выбираются из набора $\pm (15°, 30°, 45°)$.

Табл. 3.

Параметры конечных кластеров вещества
и внутренних границ в объёме тела.

| Параметры | Фиксированный диаметр | Набор диаметров |
|---|---|---|
| Мощность НКК ВГ | $(900 \pm 100) \cdot 10^{-6}$ (11.1%) | $(400 \pm 40) \cdot 10^{-6}$ (10.0%) |
| Радиус гирации НКК ВГ | $19.43 \pm 1.99$ (10.2%) | $21.28 \pm 2.37$ (11.1%) |
| Радиус-вектор центра масс НККЧ | $39.34 \pm 4.03$ (10.2%) | $36.86 \pm 3.83$ (10.4%) |
| Отношение между массой НККЧ и радиусом гирации $b = S/R^D$ | $14.64 \pm 1.54$ (10.5%) | $12.76 \pm 1.57$ (12.3%) |
| Степень анизотропии НККЧ $A_{xy}$ | $5.57 \pm 0.73$ (13.1 %) | $3.89 \pm 0.41$ (10.5 %) |
| Степень анизотропии НККЧ $A_{xz}$ | $5.42 \pm 0.69$ (12.7 %) | $4.31 \pm 0.19$ (4.4 %) |
| Степень анизотропии НККЧ $A_{yz}$ | $26.41 \pm 1.80$ (6.8 %) | $1.93 \pm 0.10$ (5.2 %) |
| Средняя корреляционная длина кластеров частиц | $38.58 \pm 3.72$ (9.6%) | $24.39 \pm 2.24$ (9.2%) |
| Индекс роста мощности КК частиц | $2.91 \pm 0.26$ (8.9%) | $3.11 \pm 0.36$ (11.6%) |
| Среднее значение радиуса гирации КК частиц | $20.42 \pm 2.44$ (11.9%) | $12.61 \pm 1.25$ (9.9%) |

Единичные треки считаются соединенными, если у них есть хотя бы одна общая точка или расстояние между ними равно некоторому минимальному расстоянию (управляющий параметр). Начальную координату трека в объёме куба, его длину и ориентационные углы определяет генератор случайных чисел с равномерным распределением.

В модельных экспериментах изучались различные режимы образования кластеров, определены условия получения бесконечных кластеров различной структуры, и рассчитаны описанные во втором разделе статьи параметры.

Результаты показали, что для степени заполнения объема тела внутренними границами характерны следующие закономерности:

– для больших треков фиксированной длины увеличение максимального

угла отклонения приводит к уменьшению фрактальной размерности перколяционного кластера, для малых длин треков наблюдается её рост;

– при фиксированном максимальном угле отклонения треков для больших углов с увеличение длины фрактальная размерность уменьшается, для малых – наблюдается рост фрактальной размерности;

– при фиксированной длине трека с увеличением максимального угла отклонения возрастет величина мощности перколяционного кластера границ;

– при фиксированном значении максимального угла: при больших углах с увеличением длины трека возрастет величина мощности кластера, при малых – зависимость обратная;

– при фиксированной длине трека с увеличением максимального угла отклонения значение перколяционного порога уменьшается;

– для больших углов при фиксированном значении максимального угла величина перколяционного порога растет с увеличением длины трека, при малых – обратная зависимость.

На рис. 6 – пример перколяционного кластера, полученного в модели.

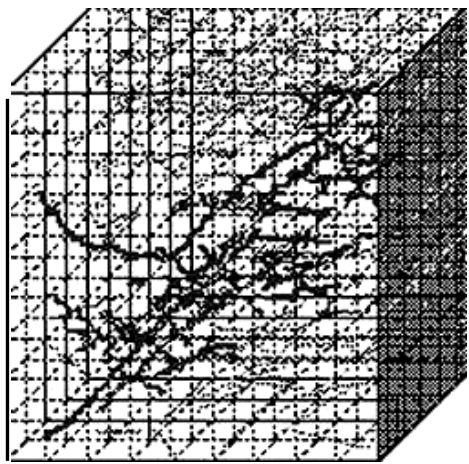

Рис. 6. Вид кластерной системы треков.

## 4. ПЕРКОЛЯЦИЯ НА ФРАКТАЛЬНОМ ПОЛЕ

Рассмотрим континуальную перколяционную задачу на ковре Серпинского с гибридной разветвленностью [11].

Ковер Серпинского является, как известно, двумерным аналогом канторовского множества исключенных средних [12]. Он может быть построен по простому алгоритму: каждая сторона квадрата единичной площади делится на три равные части; отрезки прямых, проходящие через точки деления параллельно сторонам, создают девять малых квадратов, центральный – извлекается. Процедура повторяется бесконечное число раз на каждом из восьми оставшихся; получившееся множество – регулярный фрактал с размерностью самоподобия $D = \ln 8/\ln 3 = 1,892789\ldots$ [12].

Ячейки ковра Серпинского считаются соединенными, если соприкасаются сторонами; другими словами, ковер Серпинского обладает бесконечной разветвленностью, то есть задача разделения его на части может быть решена удалением бесконечного (счетного) множества точек. Параметры перколяционного перехода на ковре Серпинского изучены в [13]. В ковре Серпинского с гибридной разветвленностью соединенными считаются клетки либо соприкасающиеся сторонами, либо имеющие общую вершину. Понятно, что модификация правил образования связности приводит к изменению перколяционных параметров бесконечного кластера ячеек ковра [11].

По описанному алгоритму разделим любую ячейку гибридного ковра Серпинского произвольного поколения на 9 клеток и удалим среднюю. Определим вероятность $p'$ принадлежности ячейки перколяционному кластеру на ковре, т.е. вероятность того, что через ячейку можно «протечь» по составляющим ее клеткам, каждая из которых входит в бесконечный кластер с вероятностью $p$. Так как ренормгрупповое преобразование [14] должно в нашем случае отражать факт наличия связности, количество подходящих комбинаций в расположении клеток в ячейке будет меньше комбинаторного. С учетом этого ренорм-преобразование для ковра с гибридной разветвленностью имеет вид

$$p' = R(p) = p^8 + 8p^7(1-p) + 27p^6(1-p)^2 +$$
$$+ 44p^5(1-p)^3 + 38p^4(1-p)^4 + 8p^3(1-p)^5,$$

с нетривиальной неподвижной точкой $p_c = 0.5093$, определяющей порог протекания [11].

Индекс длины корреляции перколяционной системы, может быть найден из соотношения $\nu = \ln b / \ln \lambda = 1.801$, где $b = 3$ – количество клеток вдоль стороны ячейки, $\lambda = (dR/dp)|p = p_c$. Критический показатель параметра порядка $\beta$ определяется из равенства $D = d - \beta/\nu$, где аппроксимацией размерности $D$ перколяционного кластера служит размерность ковра Серпинского; при размерности пространства $d = 2$ величина $\beta = 0,193$. Другие критические показатели могут быть определены из системы равенств двухпоказательного скейлинга [1]: индекс средней длины конечного кластера $\gamma = \nu d - 2\beta = 3,216$; критический показатель аналога теплоемкости $\alpha = 2 - \nu d = -1,602$; определяющий наибольший размер конечных кластеров индекс $\Delta = \nu d - \beta = 1,809$.

## 5. СИСТЕМА ИМИТАЦИОННОГО МОДЕЛИРОВАНИЯ АЛГОРИТМОВ ПОСТРОЕНИЯ И ИССЛЕДОВАНИЯ КЛАСТЕРНЫХ СТРУКТУР В МАТЕРИАЛАХ

Один из эффективных методов исследования кластерных структур базируется на алгоритмах мелкозернистого распараллеливания [15-17]. Суть алгоритмов составляет отыскание таких трансформаций исходной задачи, ее постановки, содержательного или аналитического описания, которые позволяют свести процесс решения к выполнению достаточно примитивных, пространственно распределенных и параллельных массовых вычислительных процессов. Примеры реализаций метода – ассоциативные алгоритмы решения числовых и нечисловых задач, конвейерных (систолических) и нейронных алгоритмов, конструирования клеточных автоматов, и другие [15,16]. Привлекательность подхода объясняется возможностью отыскания лучших (например, по временным характеристикам) алгоритмов решения сложных, либо громоздких и трудоемких задач. Кроме того, он служит источником методов распараллеливания алгоритмов решения задач в современных многопроцессорных вычислительных системах, и его практическая значимость определяется тем, что многие и реальные, и гипотетические компьютерные спецпроцессоры являются устройствами с мелкозернистым (клеточным) параллелизмом [15, 16].

Авторами [17] решается задача создания программного комплекса, который обеспечит конструирование алгоритмов для моделирования кластерных систем (включая перколяционные) в материалах, и позволит существенно расширить возможности их изучения. Помимо решения традиционных задач перколяционной теории, в комплексе предусмотрено:
- исследование кластеров в конфигурационных пространствах свойств;
- создание иерархии подмассивов произвольных размерностей с определяемыми радиусами взаимовлияния, направлениями действия, возможностями трансформации и другими;
- использование около трех десятков параметров для описания структуры и свойств кластеров (анизотропия, упорядоченность, связность, разветвленность, мощность, лакунарность и другие) [1, 2, 11, 12, 18-20];
- применение базирующегося на теории детерминированного хаоса [21-23] аппарата исследования особенностей эволюции кластерных систем, изучения их свойств и возможностей управления ими;
- визуализация кластерных систем и эволюционных траекторий [24].

## ЗАКЛЮЧЕНИЕ

Представленная группа моделей – апробированных и формирующихся – предназначена для решения круга задач материаловедения композитов, в которых доминирующую роль в формировании свойств играет наличие связных областей и их структура, а значит, и для решения разнообразных задач классической теории протекания.

Модель, базирующаяся на алгоритмах мелкозернистого распараллеливания, также ориентирована на перколяционные задачи в традиционных постановках. Однако ее особенности, связанные, в первую очередь, с возможностями сверхсложной организации структуры модельного объекта, могут привести и к новым постановкам, и к оригинальным результатам.

## ПРИЛОЖЕНИЕ
### Формулы для расчета параметров кластеров

При написании формул использованы обозначения:

$N$ – полное число элементов; $\langle N_s \rangle$ – среднее число кластеров размера $s$;

$s$ – количество элементов в кластере (размер кластера);

$p = \dfrac{N_{заполн.}}{N}$ – вероятность, что выбранный элемент принадлежит какому-либо кластеру;

$n_s(p) = \dfrac{\langle N_s \rangle}{N}$ – доля элементов, принадлежащих кластерам размером $s$;

$\sum_s s \cdot n_s(p)$ – доля элементов, занятых всеми конечными кластерами;

$\omega_s = \dfrac{s \cdot n_s}{\sum_s s \cdot n_s}$ – вероятность, что элемент принадлежит кластеру размером $s$;

$L^2 \cdot n_s$ – количество кластеров в квадрате $L \times L$, содержащих $s$ элементов;

$$\langle s \rangle = \sum_s \omega_s \cdot s = \frac{\sum_s s^2 \cdot n_s}{\sum_s s \cdot n_s}$$ – средняя масса конечного кластера;

$$\langle r \rangle = \frac{1}{s} \sum_s r_i$$ – радиус-вектор центра масс кластера;

$$R^2(s) = \frac{1}{2 \cdot s^2} \sum_{i,j=1}^{s} \left[ (x_{1,i} - x_{1,j})^2 + (x_{2,i} - x_{2,j})^2 \right]$$ – квадрат радиуса гирации для изотропных кластеров;

$$R^2(s) = \frac{1}{s} \sum_{i,j=1}^{s} \left[ (x_{m,i} - x_{m,j})(x_{n,i} - x_{n,j}) \right],$$ где $m, n = 1, 2$ – квадрат радиуса гирации для анизотропных кластеров;

$$A = \frac{R_{11}}{R_{22}}$$ – степень анизотропии;

$s \sim R^D$ – связь между количеством элементов в кластере и радиусом гирации, $D$ – фрактальная размерность;

$s \sim L_s^D$ – $L$ – ребро минимальной ячейки, в которую помещается кластер;

$\langle s \rangle \sim L^{2D-d}$ – средняя масса кластера;

$$n_s \sim s^{-\tau} \quad \tau = \frac{d+D}{D} = 1 + \frac{d}{D}$$ – индекс роста мощности;

$$\langle R^2(s) \rangle_L = \frac{\sum_s \langle R^2(s) \rangle \cdot s \cdot n_s}{\sum_s s \cdot n_s} \sim L^{2-(d-D)}$$ – среднее значение радиуса гирации;

$$\xi^2 = \frac{2 \sum_s R^2(s) s^2 n_s}{\sum_s s^2 n_s}$$ – корреляционная длина (длина связности);

$\xi(p) \sim |p - p_c|^{-\nu}$, $\nu$ – индекс длины корреляции;

$$P_\infty(p) = p - \sum_s s \cdot n_s(p) = \frac{N_{беск}}{N}$$ – мощность бесконечного кластера;

$\langle s^2 \rangle - \langle s \rangle^2 \sim L_s^{2D}$ – лакунарность.

## ЛИТЕРАТУРА